\documentclass[notitlepage,
superscriptaddress,
reprint,
showpacs,
preprintnumbers,
nofootinbib,
amsmath,amssymb, 
aps,
prd,
longbibliography
]{revtex4-1}

\usepackage{cancel}
\usepackage{accents}
\usepackage{mciteplus,slashed}
\usepackage{amssymb,cancel,amsmath,relsize}
\usepackage{mathrsfs} 
\usepackage{dcolumn}
\usepackage{bm}
\usepackage[caption=false]{subfig}
\usepackage{appendix}
\usepackage{physics}
\usepackage{feynmp-auto}
\usepackage[T1]{fontenc}	
\usepackage{csvsimple}
\usepackage{hyperref}
\usepackage[section]{placeins}
\usepackage[capitalise]{cleveref}
\usepackage{booktabs}
\usepackage{graphicx}
\usepackage{mathrsfs}
\usepackage[utf8]{inputenc}
\usepackage[T1]{fontenc}
\graphicspath{{Figures/}}

\hypersetup{colorlinks,citecolor= blue,linkcolor= blue, urlcolor=blue}

\usepackage[dvipsnames]{xcolor}
\usepackage[normalem]{ulem}

\usepackage{fontawesome} 

\def\e{\mathrm{e}}

\def\EfCSDA{E_f}
\def\afte{AFTE }
\DeclareMathAlphabet{\mathpzc}{OT1}{pzc}{m}{it}

\begin{document}

\preprint{FERMILAB-PUB-21-617-T} 
\preprint{IPMU21-0081}
\preprint{TTP21-051}

\title{Monopoles From an Atmospheric Fixed Target Experiment}

\author{Syuhei Iguro}
\email{igurosyuhei@gmail.com}
\affiliation{Institute for Theoretical Particle Physics (TTP), Karlsruhe Institute of Technology (KIT),
Engesserstra{\ss}e 7, 76131 Karlsruhe, Germany}
\affiliation{Institute for Astroparticle Physics (IAP),
Karlsruhe Institute of Technology (KIT), 
Hermann-von-Helmholtz-Platz 1, 76344 Eggenstein-Leopoldshafen, Germany}

\author{Ryan Plestid}
\email{rpl225@uky.edu}
\affiliation{Department of Physics and Astronomy, University of Kentucky,  Lexington, KY 40506, USA}
\affiliation{Theoretical Physics Department, Fermilab, Batavia, IL 60510,USA}

\author{Volodymyr Takhistov}
\email{volodymyr.takhistov@ipmu.jp}
\affiliation{Kavli Institute for the Physics and Mathematics of the Universe (WPI), The University of Tokyo Institutes for Advanced Study, The University of Tokyo, Kashiwa 277--8583, Japan}

\date{\today}

\begin{abstract}
Magnetic monopoles have a long history of theoretical predictions and experimental searches, carrying direct implications for fundamental concepts such as electric charge quantization.~We analyze in detail for the first time magnetic monopole production from collisions of cosmic rays bombarding the atmosphere. This source of monopoles is independent of cosmology, has been active throughout Earth's history, and supplies an irreducible monopole flux for all terrestrial experiments. Using results for robust atmospheric fixed target experiment flux of monopoles, we systematically establish direct comparisons of previous ambient monopole searches with monopole searches at particle colliders and set leading limits on magnetic monopole production in the $\sim 5-100$~TeV mass-range.
\end{abstract}

\maketitle 
  
\textbf{\textit{Introduction} -- } The existence of magnetic monopoles would symmetrize Maxwell's equations of electromagnetism and explain the observed quantization of the fundamental electric charge $e$, as demonstrated in seminal work by Dirac in 1931~\cite{Dirac:1931kp}. The charge quantization condition of $e g = n/2$, in natural units $c = \hslash = 1$ and integer $n$, establishes an elementary Dirac magnetic charge of $g_D \simeq 68.5 e$. More so, monopoles naturally appear in the context of Grand Unified Theories (GUTs) of unification of forces~\cite{Polyakov:1974ek,tHooft:1974kcl}. Despite decades of searches, monopoles remain elusive and constitute a fundamental target of interest for exploration beyond the Standard Model.

Magnetic monopoles have been historically probed through a variety of effects~\cite{Mavromatos:2020gwk}. The searches include catalysis of proton decay (``Callan-Rubakov effect'')~\cite{Rubakov:1981rg,Rubakov:1983sy,Callan:1982ac,Super-Kamiokande:2012tld}, modification of galactic magnetic fields (``Parker bound'')~\cite{Parker:1970xv}, Cherenkov radiation~\cite{IceCube:2015agw,BAIKAL:2007kno} and ionization deposits due to monopoles accelerated in cosmic magnetic fields and contributing to cosmic radiation~\cite{MACRO:2002jdv}. 

The majority of monopole searches have relied on an abundance of cosmological monopoles, as produced in the early Universe via the Kibble-Zurek mechanism~\cite{Kibble:1976sj,Zurek:1985qw}. However, this is highly sensitive to model details, such as mass of the monopoles and the scale of cosmic inflation expansion that could significantly dilute any pre-existing monopoles. This results in a large uncertainty in the interpretation of monopole searches.

While many of the previous studies have focused on ultra-heavy GUT-scale monopoles (i.e.\ $\sim 10^{16}$~GeV masses), scenarios exist with monopole masses $M \ll 10^{16}$ GeV, which are often called intermediate mass monopoles~\cite{Kephart:2017esj}.
Recently, reinvigorated interest in monopole searches has been fueled by the identification of scenarios with viable electroweak-scale monopoles ~\cite{Cho:1996qd,Cho:2013vba,Ellis:2016glu,Arunasalam:2017eyu,Arai:2018uoy,Ellis:2017edi}. Sensitive searches of TeV-scale monopoles have been carried out at the Large Hadron Collider's (LHC) ATLAS~\cite{ATLAS:2019wkg} and MoEDAL experiments~\cite{MoEDAL:2019ort}.

In this work we explore for the first time monopole production from collisions of cosmic rays bombarding the atmosphere. Historically, atmospheric cosmic ray collisions have been employed as a flagship production site for neutrino studies, leading to the discovery of neutrino oscillations~\cite{Super-Kamiokande:1998kpq}. The resulting monopole flux from the ``atmospheric fixed target experiment'' (\afte) is independent of cosmological uncertainties, and this source has been active for billions of years throughout Earth's history. More importantly, this robust source of monopoles is universal and potentially accessible to all terrestrial experiments. This opens a new window for magnetic monopoles searches and allows us to establish the first direct comparison between constraints from colliders and historic searches based on an ambient cosmic monopole abundance.
 
\begin{figure}[tb]
\includegraphics[trim={35mm 10mm 20mm 0mm},clip,width=.60\textwidth]{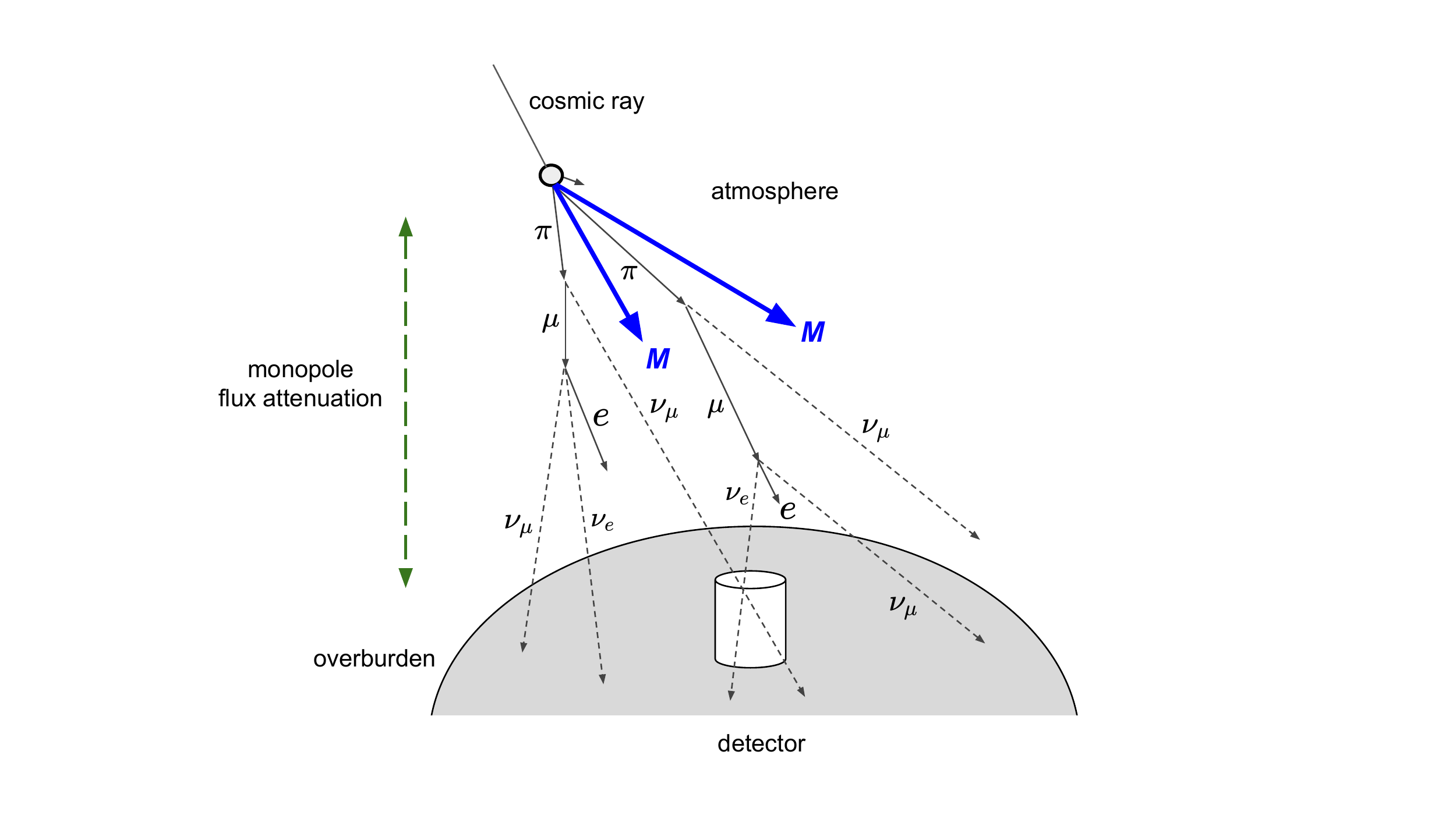}
\caption{A schematic of the magnetic monopole $(M)$ production from atmospheric cosmic ray collisions. \label{fig:atmcmonopole}
}
\end{figure}

\textbf{\textit{Monopole production} -- }
Collisions between incoming isotropic cosmic ray flux and the atmosphere results in copious production of particles from the model spectrum~\cite{ParticleDataGroup:2020ssz}, as depicted on Fig.~\ref{fig:atmcmonopole}. Focusing on the dominant proton $p$ constituents, \afte is primarily a source of proton-proton ($pp$) collisions. Unlike, conventional collider experiments that operate at a fixed energy, the LHC being $\sim 10$ TeV-scale, the cosmic ray flux allows for the exploration of new physics with \afte  over a broad energy spectrum reaching monopole masses as large as $\sim 10^6$ GeV.

For detailed analysis of \afte  monopole $M$ flux production we perform Monte Carlo simulations of $pp~\rightarrow~ M\overline{M}$ processes, drawing on the methodology of the LHC MoEDAL experiment~\cite{MoEDAL:2017vhz,MoEDAL:2019ort}.~In particular, we employ {\sc\small MadGraph}5 (MG5) version 3.1.0~\cite{Alwall:2014hca} simulation tools with NNPDF31luxQED parton distribution functions~\cite{Bertone:2017bme} and input UFO files of Ref.~\cite{Baines:2018ltl} for each incident proton energy in the COM frame. This procedure allows for an ``apples-to-apples'' comparison between cosmic ray observations and collider searches. 

\begin{figure}[tb]
\includegraphics[trim={45mm 30mm 40mm 30mm},clip,width=.49\textwidth]{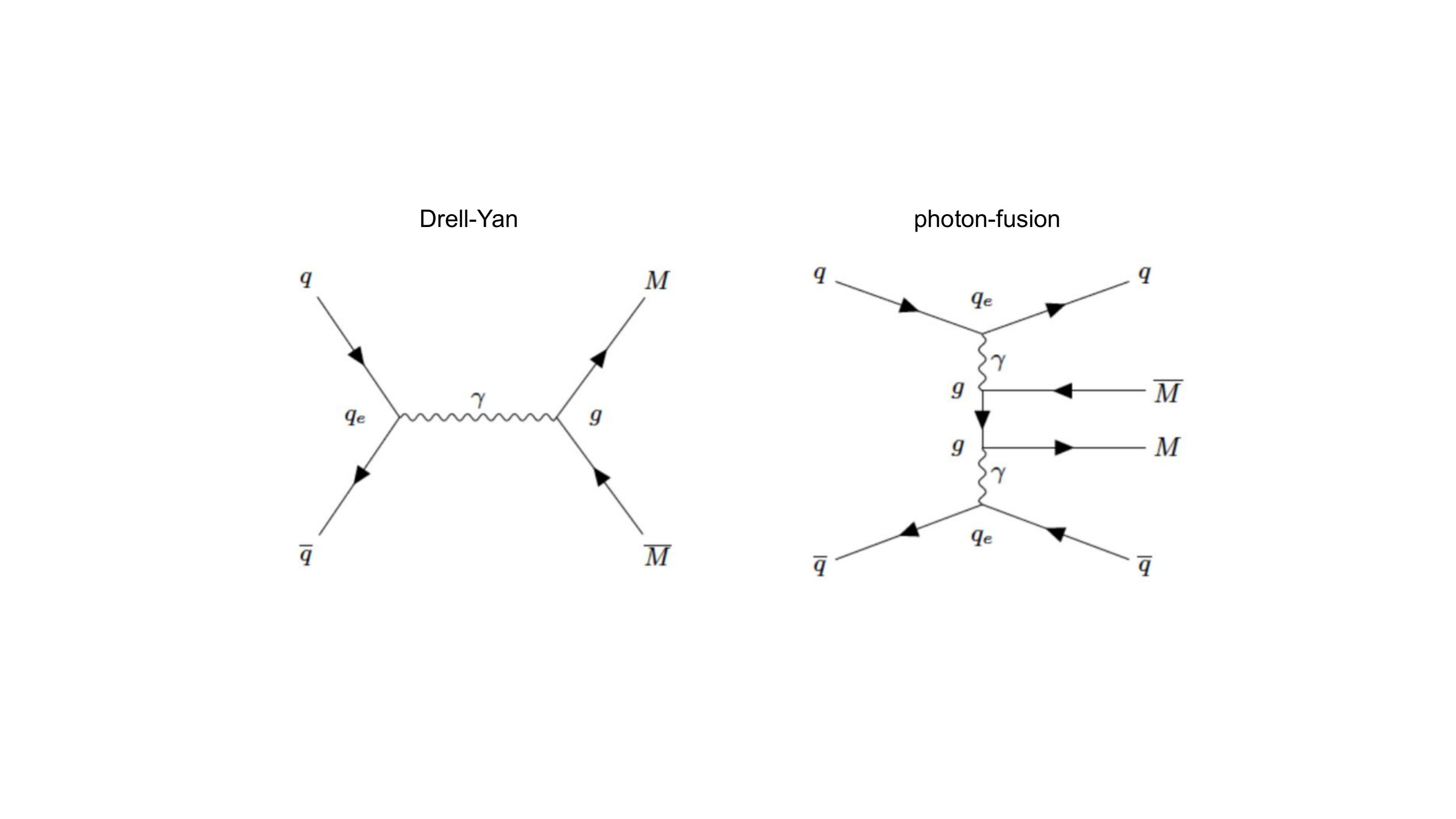}
\caption{
Diagrams for Drell-Yan and photon-fusion monopole production processes in $pp$ collisions.  \label{fig:monopoleproc}
}
\end{figure}

We model monopole production in hadronic collisions by tree-level Feynman diagrams appropriate for an elementary charged particle~\cite{Baines:2018ltl}, as employed in LHC searches~\cite{MoEDAL:2017vhz,MoEDAL:2019ort}. In particular, as depicted on Fig.~\ref{fig:monopoleproc}, we consider the traditional Drell-Yan (DY) production\footnote{Recently, symmetry arguments within certain classes of theories have been put forth that question Drell-Yan monopole production~\cite{Terning:2020dzg}. In this work we remain agnostic about this traditional monopole channel and include it for direct comparison with existing searches.}
for magnetic monopoles via quark-pair annihilation through a virtual photon $q\bar{q} \rightarrow \gamma^* \rightarrow M\overline{M}$, as well as photon-fusion (PF) $\gamma^*\gamma^* \rightarrow M\overline{M}$. We find numerically that photon fusion always dominates over Drell-Yan production. Our methodology is in contrast to some other new physics searches with \afte\footnote{Ultra-high energy neutrino collisions have been also previously considered, e.g.~\cite{Anchordoqui:2001cg,Emparan:2001kf,Feng:2001ib,Jho:2018dvt}.}, which, in analogy with atmospheric neutrino studies~\cite{Super-Kamiokande:1998kpq}, have been primarily targeting meson decays~\cite{Coloma:2019htx,Plestid:2020kdm,ArguellesDelgado:2021lek,Candia:2021bsl}. Monopoles are also distinct in that they are strongly coupled. Hence, monopoles may have large production cross sections that could allow a non-negligible flux even for PeV-scale cosmic rays. 
 
A monopole pair production requires that the square of the COM energy is $s\geq 4 M^2$. This necessarily leads to highly boosted kinematics in the lab frame. The boost factor relating the lab-frame and COM frame is $\gamma_{\rm cm}= \sqrt{s}/(2 m_p)$, where $m_p$ is the proton mass, which leads to $\gamma_{\rm cm}\geq M/m_p$. 
Since the cosmic ray flux falls rapidly with increasing proton kinetic energy, it is expected that the majority of monopoles are produced near threshold. The typical collision lab-frame energy is  $\langle E_M \rangle \sim M^2/m_p$. 
  
For simplicity, we focus on a spin-half\footnote{We note that the resultant cross-section is $\sim10$ times larger than that of spin-zero model but $\sim10$ times smaller than that of spin-one model~\cite{MoEDAL:2019ort}}, $g_D = 1$ and velocity ($\beta$-)independent monopole model\footnote{In order to discuss a wider range of theoretical models, some studies have also advocated for considering a $\beta$-dependent   coupling $g \beta = g \sqrt{1 - 4 M^2/s^2}$~(e.g.~\cite{Epele:2012jn}). The $\beta$-dependence of the model can suppress the cross section by a factor of 2.}. We have confirmed that qualitatively our comparison between different searches will not be significantly impacted by this choice. Our analysis can be readily extended to other possibilities. The resultant monopoles are then boosted to the lab frame, with $    E_{\rm lab} = \gamma_{\rm com} E_{\rm com} + \gamma_{\rm com} \beta_{\rm com} P_{\rm com}\cos\theta$ where $\cos\theta$ is the angle of the monopole momentum relative to the proton momentum. 
 
From each simulation we obtain an overall $pp$ interaction cross section $\sigma(pp\rightarrow M\overline{M})(s)$, which is then used for comparison with data.  Since lab frame distributions are primarily dictated by kinematics, a significant uncertainty in modelling the monopole interactions here (using tree-level Feynman diagrams to model strongly coupled theory with monopoles) is an overall normalization. We employ our simulation results as a model of the monopoles' kinematic distribution, but allow the overall normalization of the cross-section to be a free parameter 
\[\sigma(pp\rightarrow M\overline{M})= \kappa \times \sigma_{\rm sim}~,
\]
where $\sigma_{\rm sim}$ is the simulation output cross-section and $\kappa$ is a constant.  

The outlined procedure allows us to consistently and directly compare different collider monopole searches by constraining $\kappa$ for different monopole masses, accomplished by taking the ratio of the limited cross section and the simulation predictions at a particular energy. We choose a reference cross section defined at $\sqrt{s} =  4 M$, twice the threshold production energy. 

\begin{figure}[t]
\centering\includegraphics[width=0.95\linewidth]{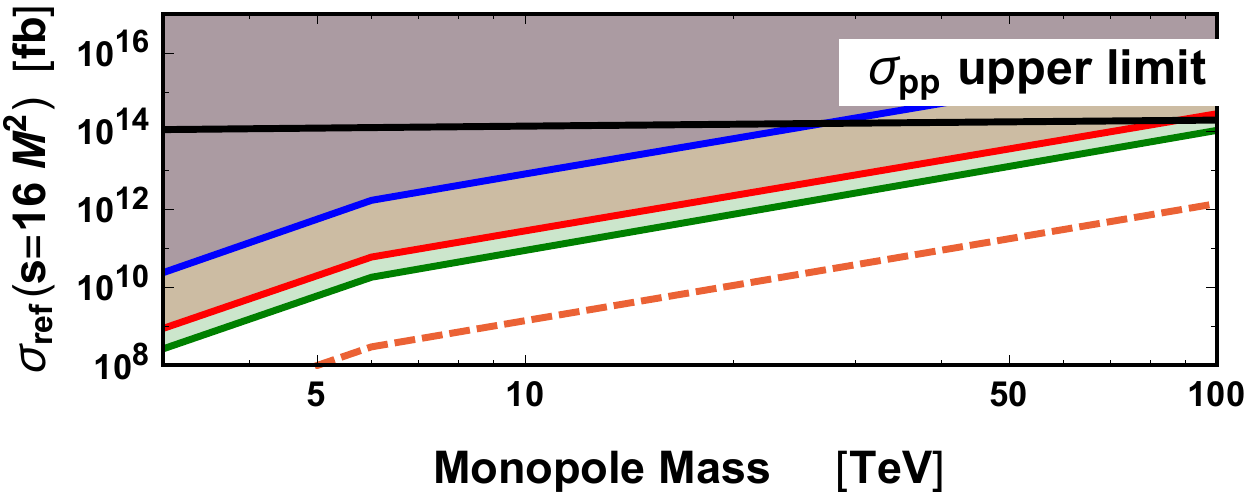}
\includegraphics[width=0.95\linewidth]{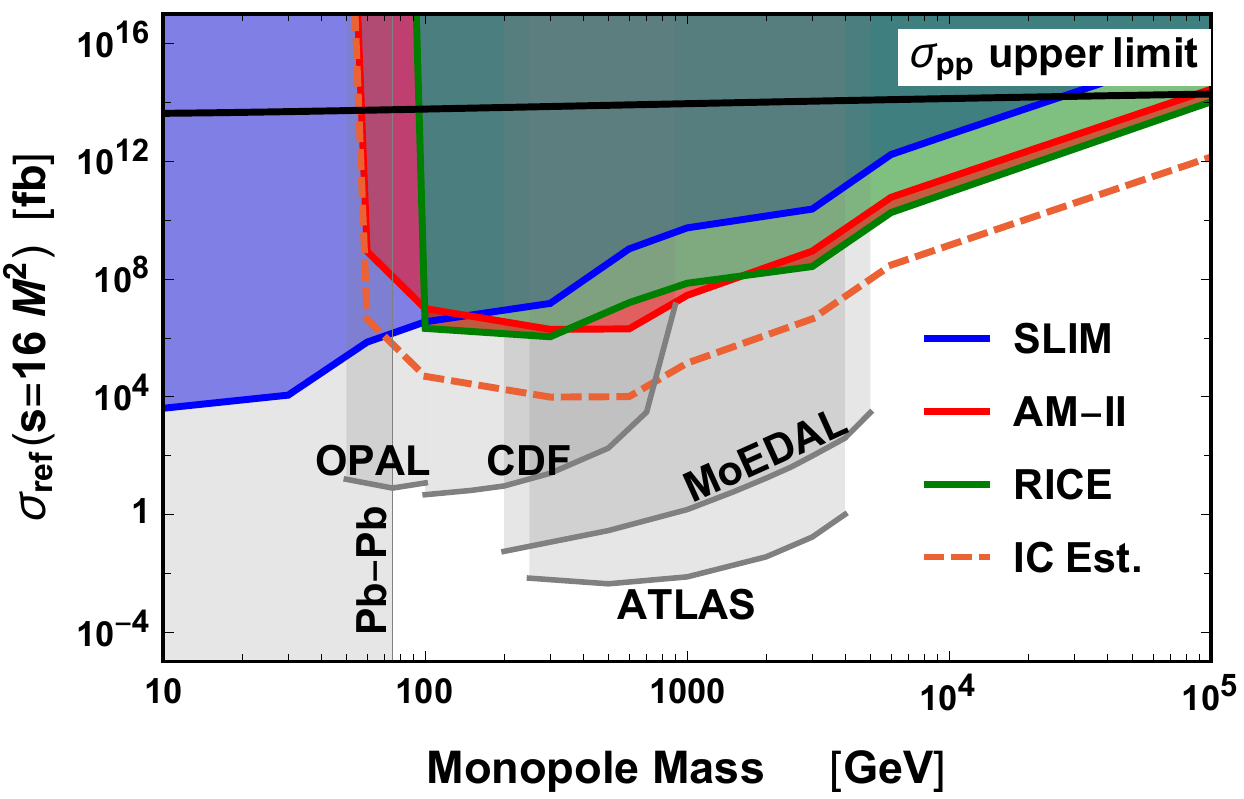}
\caption{\label{fig:limits} Comparison of novel monopole limits (shaded regions) from cosmic ray atmospheric collisions derived in this work using historic data from SLIM, AMANDA-II (AM-II) and RICE experiments. Also displayed is a crude projection for a down-going monopole search in IceCube experiment (IC Est.) defined by multiplying the sensitivity from AMANDA-II by a factor of 200 (see ~\cite{IceCube:2021eye}). Comparison is systematically achieved using the common reference cross section for $pp\rightarrow M\overline{M}$ defined by $\kappa \times \sigma_{\rm sim}(s=16 M^2)$, where the normalization $\kappa$ is found by comparing the simulation predictions to the derived constraints for each target experiment. Existing limits for collider monopole searches by OPAL~\cite{OPAL:2007eyf}, CDF~\cite{CDF:2005cvf}, MoEDAL~\cite{MoEDAL:2019ort}, and ATLAS~\cite{ATLAS:2019wkg} experiments are shown. We have further excluded monopole masses less than $75$ GeV due to constraints from Pb-Pb collisions that rely on the calculable Schwinger pair production cross section~\cite{Acharya:2021ckc}. Also displayed is the total $pp\rightarrow X$ cross section, as parameterized by the COMPETE collaboration~\cite{COMPETE:2002jcr}, which sets an upper limit on the allowed $pp\rightarrow M\overline{M}$ cross section.}
\end{figure}

As a demonstrative example, taking a monopole mass of 150 GeV, we compare the constraints for the CDF experiment $\sigma_{\rm CDF} \lesssim 0.4$ pb at $\sqrt{s}_{p\bar{p}} =1.96$ GeV \cite{CDF:2005cvf} to the prediction from our simulations for $p\bar{p}$ collisions $\sigma_{\rm sim}(\sqrt{s}=1.96~{\rm GeV}) = 2.6~{\rm pb}$. This gives $\kappa=\sigma_{\rm sim}/\sigma_{\rm CDF}\approx 6.5$. Next, we compute the reference cross section at $\sqrt{s}=4 M$, i.e.\ $2\times$ larger than threshold. This results in $\sigma_{\rm ref}^{[\rm{CDF}]}(M=150~{\rm GeV}) = 6.5 \times \sigma_{\rm sim} (\sqrt{s}=600~{\rm GeV})$. The same procedure is used for all of the experiments in this work that are then plotted in terms of their reference cross sections in \cref{fig:limits}. 

\begin{figure}[t]
\centering
\includegraphics[scale=0.24]{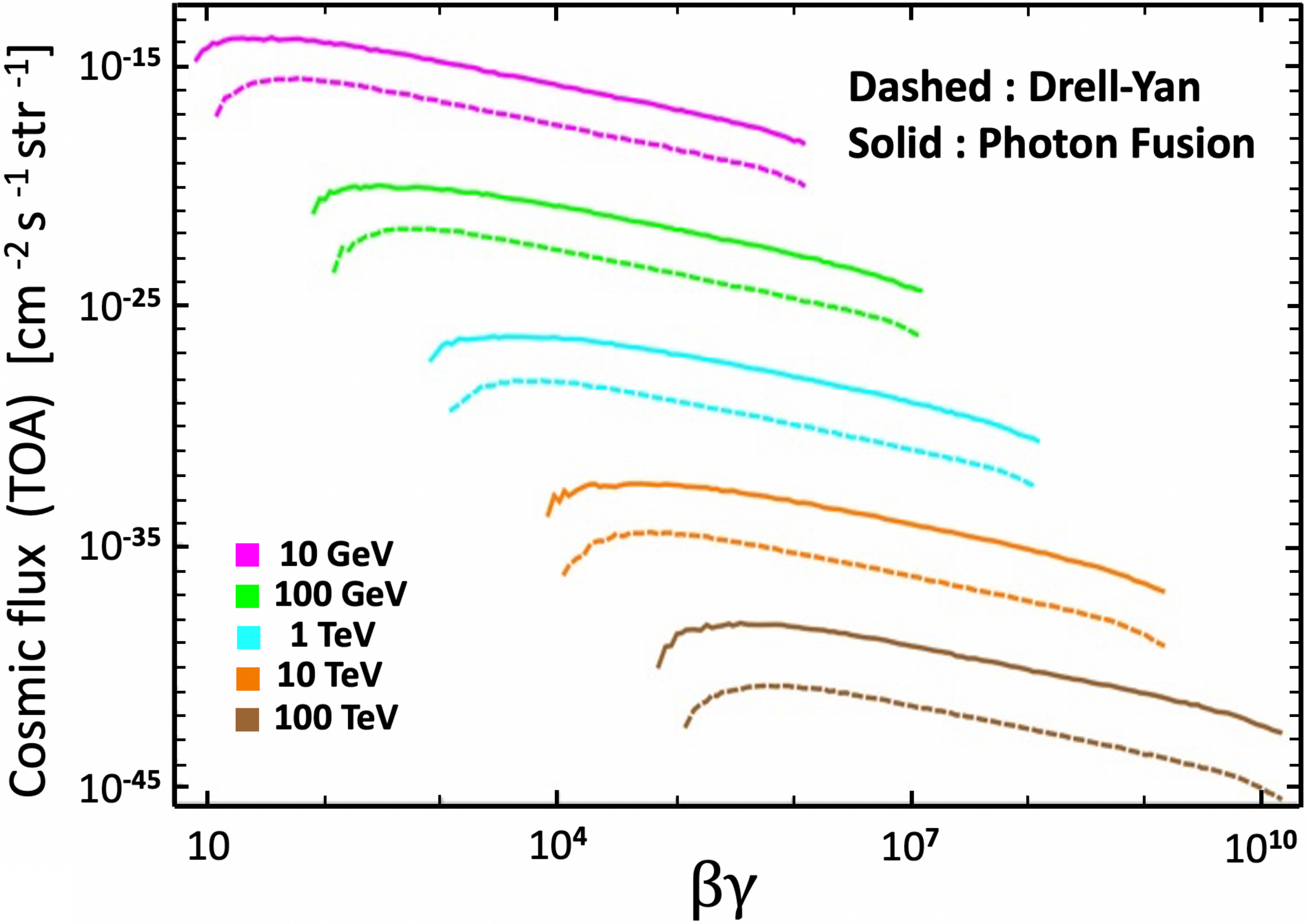}
\caption{
Flux intensity of monopoles produced in the thick-target approximation (see text) from the top-of-atmosphere cosmic ray collisions as predicted by simulations and considering cross-section normalization of $\kappa=1$. The net produced monopole number is two times the event number ($pp \rightarrow M\overline{M}$). Different monopole masses as well as Drell-Yan and photon-fusion production processes are shown. \label{fig:monopoleprod}
}
\end{figure}

Considering that all incoming cosmic protons are eventually absorbed in the atmosphere,
the cross-section calculated with simulations is convoluted with the inelastic cross section\footnote{We note that at masses that are larger than the kinematic reach of the LHC, the measurements for $pp\rightarrow X$ have increased uncertainty, stemming from cosmic ray data which has uncertainties on the order of $\sim 30\%$ (see e.g.\ \cite{ParticleDataGroup:2020ssz}).
We note that $\sigma(\sqrt{s})$ varies between $\sim 40$ mb and $\sim 100$ mb as $\sqrt{s}$ ranges from 10 GeV to 100 TeV.} $\sigma_{\rm inel}$ and cosmic proton density~\cite{ParticleDataGroup:2020ssz}. 
In \cref{fig:monopoleprod} we show the resulting weighted \afte  monopole flux event number for $pp\to M\overline{M}$ as a function of relativistic $\beta\gamma$.  

The behavior of the monopole flux can be approximately understood as follows. Our simulation found that the parton level cross section is proportional to $\sqrt{E_M}$.
The cosmic proton number density rapidly decays as a function of the energy, $\propto E_M^{-3}$~\cite{ParticleDataGroup:2020ssz}.
Hence, assuming production near threshold, the resulting monopole flux approximately behaves as $\propto M^{-5}$, since it is proportional to cross section multiplied by the cosmic ray density.
 
\textbf{\textit{Flux attenuation} -- } As the monopoles propagate through a medium, their interactions result in attenuation of the flux. We note that the number of monopoles is conserved, however their energy is modified. We outline the treatment of AFTE monopole flux attenuation, and provide additional computational details in Supplemental Material.

The procedure discussed thus far yields \afte  flux of monopoles that would be produced per proton in a finite ``thick target'' at the top of the atmosphere. Effects of attenuation of the incoming cosmic ray flux can be understood through proton mean free path $\lambda=1/[n(s) \sigma_{\rm inel}(E_p)]$, where $E_p$ is the proton energy and $n(s)$ is the density profile of air taken from the global reference atmospheric model~\cite{Picone:2002xxx}. 
 
The energy losses of monopoles produced at a particular height that propagate through a medium of density $n(x)$ towards a given detector are readily accounted for through stopping power per unit length $\dd E / \dd x = f(E) n(x)$, where $f(E)$ is the attenuation function. Effects responsible for energy deposit of monopoles traversing a medium depend on $\beta\gamma$ and velocity. We focus on monopoles with $\beta\gamma\gtrsim 0.03$, relevant for the detectors of interest. Hence, we can reliably estimate the stopping power $\dd E/ \dd x$ of monopoles passing through matter using the standard Bethe-Bloch-like formula for ionization losses~\cite{ParticleDataGroup:2020ssz}, applicable for $0.03 \lesssim \beta \gamma \lesssim 10^4 $ kinematic regimes. 
At still higher energies monopole energy losses are dominated by photonuclear processes~\cite{ANITA-II:2010jck} and are expected to grow super-linearly with $\beta\gamma$ for $\beta\gamma \gtrsim 10^4$, with an approximate scaling of $\dd E/\dd x \sim \gamma^{1.2}$. This introduces an effective maximum velocity cutoff for propagating monopoles, since any ultra-relativistic monopole is rapidly decelerated until it hits the plateau of the Bethe-Bloch ionization. The breaking effect from photon-nuclear reactions ensures that any monopoles reaching the Earth's surface have $\beta\gamma \lesssim 10^4-10^5$.

\textbf{\textit{Experimental searches} -- }  Using universal persistent \afte monopole flux available for all terrestrial experiments we can re-analyze historic data from ambient monopole flux searches.

For a high-altitude experiment, the attenuation of the monopole flux by the atmosphere will set a lower bound on the mass of monopoles that can reach the detector. Given experimental sensitivity threshold to monopoles at $\beta_{\rm min}$, the detector's signal intensity is then found by appropriately integrating the resulting attenuated flux to account for monopoles produced at a particular height that will be travelling faster than the cutoff $\beta'\geq \beta_{\rm min}$ at the experiment.
For the propagation of the flux through the atmosphere, the lowest energy monopoles are most important. Thus, we approximate $\sigma_{\rm inel} = 40$ mb, which is valid for 5 GeV $\lesssim E_p \lesssim 10^4$ GeV~\cite{ParticleDataGroup:2020ssz}, in calculating the mean free path $\lambda=1/(n \sigma_{\rm inel})$. 

Deep underground experiments, located under sizable overburden, are largely insensitive to atmospheric effects. Monopoles that can penetrate the overburden will loose negligible energy while traversing the atmosphere. Hence, the intensity of monopoles arriving at the surface can be reliably approximated by the intensity at the top of the atmosphere. For detectors whose column density of overburden satisfies $\rho_\perp({\rm overburden}) \gg \rho_\perp({\rm air})$, we take into consideration the zenith-angle dependence of the intensity\footnote{The background of cosmic ray muons depends on zenith angle and so experimental cuts are necessarily more severe for down-going monopoles. Because the monopole flux is attenuated at large zenith angles there is a competition between these two effects such that an optimum zenith angle will be achieved at intermediate values $\theta_z\sim \pi/6-\pi/3$.}.

Variety of distinct historic analyses have searched for ambient astrophysical monopole flux. Existing experimental limits\footnote{We do not consider here searches focusing on GUT monopoles, e.g. Super-Kamiokande~\cite{Super-Kamiokande:2012tld}.} include those from AMANDA-II~\cite{Abbasi:2010zz}, IceCube~\cite{IceCube:2015agw}, MACRO~\cite{MACRO:2002iaq}, SLIM~\cite{Balestra:2008ps}, NOvA~\cite{NOvA:2020qpg}, ANITA-II~\cite{ANITA-II:2010jck}, and the Baikal observatory~\cite{BAIKAL:2007kno}. However, as we discuss, some limits are not applicable to \afte monopoles.

By limiting the analysis to up-going monopoles, experiments can suppress atmospheric muon backgrounds. However, implicitly such monopoles traverse the bulk of the Earth with path lengths on the order of thousands of kilometers. As this is not applicable to \afte monopoles, we do not consider such limits from IceCube~\cite{IceCube:2015agw} or Baikal~\cite{BAIKAL:2007kno} that restricted zenith angle of the incoming monopole direction to be up-going. Searches focusing on slow-moving monopoles with $\beta \gamma \lesssim 10$, such as of
MACRO~\cite{MACRO:2002jdv} and NOvA~\cite{NOvA:2020qpg}, are also ineffective for probing \afte monopoles. Further, \afte flux is also highly suppressed for ultra-relativistic monopole searches, such as of ANITA-II \cite{ANITA-II:2010jck}, as can be seen from Fig.~\ref{fig:monopoleprod}. 

The RICE underground experiment focused on detecting radio emission from in-ice monopole interactions in regimes relevant for \afte monopoles, with $\beta\gamma \gtrsim 10^7$ ~\cite{Hogan:2008sx}. At such large boosts the attenuation from the air is negligible, but not from Earth. We reinterpret RICE limits for \afte monopoles, multiplying them by an additional factor of 2 to approximately account for the absence of up-going monopoles. We employ the monopole flux limits of Ref.~\cite{Hogan:2008sx} for $\gamma=10^7$ and $\gamma=10^8$ and compare them to our simulation predictions integrated over $\gamma\in[10^{6.5},10^{7.5}]$ and $\gamma\in[10^{7.5},10^{8.5}]$, respectively.
The resulting novel limits are displayed in Fig.~\ref{fig:limits}.

Particularly favorable for \afte monopoles is the SLIM nuclear track experiment~\cite{Balestra:2008ps}, sensitive to lighter mass monopoles due to its high elevation of 5230 m above sea level.
In setting limits we use SLIM's constraint for $\beta \geq 0.03$, requiring that monopoles reaching the detector have $\beta\gamma \geq 0.03$. As the search is for purely down-going monopoles, we can employ appropriate flux intensity directly together with the bound of $\mathcal{I}(\cos\theta_z=1) \leq 1.3 \times 10^{-15}~{\rm cm}^{-2} ~{\rm s}^{-1} ~{\rm str}^{-1}$. The newly established bounds on monopoles from \afte by SLIM are superseded by collider searches at lower masses, as well as RICE and AMANDA-II at higher masses, see Fig.~\ref{fig:limits}. 

Dedicated search for down-going monopoles has been performed by the deep-ice AMANDA-II experiment~\cite{Abbasi:2010zz}. Here we take into account the zenith angle dependence of the monopole detection efficiency $\epsilon(\cos \theta_z)$ extracted from data. We employ the resulting constraints for Cherenkov emission from a $\beta=1$ monopole, and require $\beta\gamma \geq 3$ corresponding to $\beta \geq 0.95$.  The down-going monopole search of AMANDA-II imposes a cut on the zenith angle and a cut in the space of $\cos\theta_z$ and $\Sigma ADC$, a quantity related to the sum of the photomultiplier tube pulse amplitudes. We infer the efficiency as a function of $\Sigma ADC$ from Fig.\ 11  and the cut on $\Sigma ADC$ as a function of $\cos\theta_z$ from Fig.\ 12 of Ref.~\cite{Abbasi:2010zz}. The AMANDA-II bounds assume an isotropic flux of monopoles such that $\Phi \leq C\big/\int \epsilon(\cos\theta_z) \dd \cos\theta_z$. Extracting $C$ and including an appropriate zenith angle flux we then compute the cross-section normalization factor $\kappa$. The results are depicted on Fig.~\ref{fig:limits}. We find that AMANDA-II and RICE establish comparable monopole limits.

\textbf{\textit{Conclusions} -- } Magnetic monopoles are directly connected with different aspects of fundamental physics and have been historically a prominent topic of both theoretical and experimental investigations. We have analyzed for the first time monopole production from atmospheric cosmic ray collisions. This source of monopoles is not subject to cosmological uncertainties and is persistent for all terrestrial experiments. Using historic data from  RICE, AMANDA-II and SLIM experiments together with monopole flux from atmospheric cosmic ray collisions, we have established leading robust bounds on the production cross section of magnetic monopoles in the $\sim 5-100$ TeV mass-range. We project that a dedicated search from IceCube could potentially set the best limits on monopole masses larger than 5 TeV that lie beyond the reach of current colliders. 

~\newline
\textbf{\textit{Acknowledgements} -- }
We thank Mihoko Nojiri for discussion of the parton distribution functions. The work of S.I. is supported by the JSPS Core-to-Core Program (Grant No.\,JPJSCCA20200002). V.T. is supported by the World Premier International Research Center Initiative (WPI), MEXT, Japan. This work was performed in part at Aspen Center for Physics, which is supported by National Science Foundation grant PHY-1607611. R.P.\ is supported by the U.S. Department of Energy, Office of Science, Office of High Energy Physics, under Award Number DE-SC0019095. This manuscript has been authored by Fermi Research Alliance, LLC under Contract No. DE-AC02-07CH11359 with the U.S. Department of Energy, Office of Science, Office of High Energy Physics. 

\bibliographystyle{utphys28mod}

\bibliography{bibfile}

\clearpage
\onecolumngrid
\begin{center}
   \textbf{\large SUPPLEMENTAL MATERIAL \\[.1cm] ``Monopoles From an Atmospheric Fixed Target Experiment''}\\[.2cm]
  \vspace{0.05in}
  {Syuhei Iguro, Ryan Plestid, Volodymyr Takhistov}
\end{center}

\twocolumngrid
\setcounter{equation}{0}
\setcounter{figure}{0}
\setcounter{table}{0}
\setcounter{section}{0}
\setcounter{page}{1}
\makeatletter
\renewcommand{\theequation}{S\arabic{equation}}
\renewcommand{\thefigure}{S\arabic{figure}}
\renewcommand{\thetable}{S\arabic{table}}

\onecolumngrid

In this Supplemental Material, we describe additional details for computation of monopole flux attenuation and for obtaining monopole flux intensity relevant for experiments.
 
\section{Flux attenuation}
 The simulations procedure outlined in the main text establishes \afte flux of monopoles produced per proton in a finite ``thick target'' at the top of the atmosphere (TOA). Accounting for the fact that the atmosphere has an altitude-dependent density, the TOA proton flux $\mathcal{I}_p^{\rm TOA}$ must be replaced with an attenuated proton flux that depends on both the proton energy $E_p$ and height $z$ (thickness of target), 

\begin{equation}
   \label{Intensity-1}
   \mathcal{I}_p (z,E_p)= \exp\qty[-\int_z^\infty \frac{ \dd s}{ \lambda} ] ~\mathcal{I}_p^{\rm TOA}(E_p)~, 
\end{equation}
 where $\lambda=1/[n(s) \sigma_{\rm inel}(E_p)]$ is the mean free path of a proton, and $n(s)$ is the density profile of air. 
 
Monopoles produced at $z$ must then  propagate to a given detector at a height $z_0$. In the continuous slowing down approximation (CSDA) the energy of the monopole is deterministic, with an initial monopole energy $E_M$ mapped to unique final energy $E_M'=\EfCSDA$. This can be computed using energy loss per unit length stopping power $\dd E / \dd x = f(E) n(x)$ for monopoles passing through a medium of density $n(x)$ from a point $z_i$ to a point $z_f$ by solving
 \begin{equation}
    \label{CSDA}
    \int_{E_f}^{E_i} \frac{1}{f(E)} \dd E = \int_{z_f}^{z_i} ~n(x) \dd x 
 \end{equation}
 for the final energy $E_f$. The result defines $\EfCSDA(z_i,z_f, E_i)$. Since $f(E)= f(\beta\gamma)$, the attenuation for different monopole masses can be treated by a rescaling of \cref{CSDA}. 

The monopole intensity at a detector is then given by
 \begin{equation}\label{Intensity-2}
    \begin{split}
       \mathcal{I}_M (z_0, E_M')
       =& \int \dd E_p \dd z ~~T(E_M' | E_M) \\
       &\times \mathcal{I}_p (z,E_p)~n(z) \frac{\dd \sigma(E_p)}{\dd E_M} ~,
    \end{split}
 \end{equation}
where $\mathcal{I}$ is defined in Eq.~\eqref{Intensity-1},  $T(E_M'|E_M)$ is the CSDA transfer matrix $T(E_M'|E_M)  = \delta\big(E_M' - \EfCSDA(z_i,z_f, E_M)\big)$. 
The $pp$ inelastic cross section is a very slowly varying function of $E_p$ at high energies and can be approximated by a constant value. We may then factorize Eq.~\eqref{Intensity-2} into two integrals
\begin{align}
    \label{Intensity-3}
       \mathcal{I}_M (z_0, E_M') 
       =&~\int \dd E_p  ~ ~\mathcal{I}_p^{\rm TOA}(E_p) ~\frac{1}{\sigma_{\rm inel}}\frac{\dd \sigma}{\dd E_M}\\
       &~\times \qty[\int_{z_0}^\infty \dd z ~ \frac{1}{\lambda(z)}~T(E_M' | E_M)\e^{-\int_{z}^\infty \dd s/ \lambda} ]~. \nonumber
\end{align}
The top line of this formula may be interpreted as the primary flux of monopoles passing through a thick target, $\mathcal{I}_{TT}(E_M)$. 

\section{Monopole Flux Intensity for Experiments}
Given experimental sensitivity threshold to monopoles at $\beta_{\rm min}$, the detector's signal intensity is then found by integrating Eq.~\eqref{Intensity-3} from $\beta^\prime = \beta_{\rm min}$ to $\infty$. Within the CSDA, for each fixed $z$ at which a monopole is produced, there is a well defined $E_M^{[{\rm min}]}(z)$ above which monopoles will be travelling faster than the cutoff $\beta'\geq \beta_{\rm min}$ at the experiment.

Hence, the signal flux of down-going monopoles reaching the high altitude experiment $\mathcal{I}_M^{\rm high}$ is given by
\begin{equation}
    \label{high}
    \mathcal{I}_M^{\rm high} = \int_{z_{\rm exp}}^\infty \hspace{-10pt} \dd z 
    \frac{\e^{- \int_{z}^\infty \tfrac{\dd s }{\lambda(s)}}}{\lambda(z)} \hspace{1pt}\int_{E_M^{[{\rm min}]}(z)}^\infty
    \hspace{-5pt}~\dd E_M  \mathcal{I}_{TT}(E_M)~,
\end{equation}
where $z_{\rm exp}$ is the height of experimental site above sea level. This can be readily generalized to different zenith angles.
Here, we treat $\sigma_{\rm inel}$ as a function of $E_p$ in calculating $\mathcal{I}_{TT}$. 

Monopoles that can penetrate the overburden of a deep underground experiment and loose negligible energy traveling through the atmosphere can have their intensity of monopoles arriving at the surface approximated by $\mathcal{I}_{TT}(E_M)$. For detectors whose column density of overburden satisfies $\rho_\perp({\rm overburden}) \gg \rho_\perp({\rm air})$, we instead focus on the zenith-angle dependence of the intensity. 

The resulting path length that a monopole must travel through the overburden, $\ell$, for a detector with an overburden of depth $d$, is given by 
\begin{equation}
    \ell=\sqrt{\cos ^2\theta_z \! \left(R-d \right)^2+d \left(2R -d\right)}-\cos \theta_z\! \left(R-d\right),
\end{equation}
where $R$ is the Earth's radius. Using \cref{CSDA} we can take $z_i=\ell$, and fix $E_f\geq E_{\rm thr}$ for the threshold energy (or equivalently velocity $\beta_{\rm min}$) of the given experimental search. This defines  $E_M^{[{\rm min}]}(\cos\theta_z)$.  Including a zenith angle-dependent efficiency, $\epsilon(\cos\theta_z)$, the observed intensity at an underground experiment, $\mathcal{I}_M^{\rm und}$, corresponding to an integrated observed intensity $\mathcal{I}_M^{\rm obs}$ over a solid angle is given by
\begin{align}    
\label{deep} 
    \mathcal{I}_M^{\rm und} =&~ \int \dd \Omega ~\mathcal{I}_M^{\rm obs} \notag\\ 
    =&~ 
    2\pi \int  \dd\cos\theta_z ~\epsilon(\cos\theta_z) \\
    &
    \times\int_{E_M^{[{\rm min}]}(\cos\theta_z)}^\infty \hspace{-10pt}\dd E_M~~ \mathcal{I}_{TT}(E_M)~. \notag 
\end{align}

\end{document}